\documentclass[a4paper,onecolumn,11pt]{quantumarticle}
\toggletrue{@unpublished}
\pdfoutput=1
\usepackage[utf8]{inputenc}
\usepackage[english]{babel}
\usepackage[T1]{fontenc}
\usepackage{amsmath}
\usepackage{amssymb}
\usepackage{hyperref}
\usepackage{xcolor}
\usepackage{datetime}

\usepackage{tikz}
\usetikzlibrary{quantikz}
\usepackage{lipsum}
\usepackage[square,numbers]{natbib}
\usepackage[ruled,vlined]{algorithm2e}

\def\Javier{blue}
\newcommand{\figref}[1]{\figurename~\ref{#1}}

\title{Hermitian Matrix Definiteness from Quantum Phase Estimation}
\author[1]{Andr\'es G\'omez}
\affil[1]{Centro de Supercomputación de Galicia (CESGA)}
\email{agomez@cesga.es}
\author[2]{Javier Mas}
\affil[2]{  Departamento de  F\'\i sica de Part\'\i  culas, Universidade de Santiago de Compostela and  
 Instituto Galego de F\'\i sica de Altas Enerx\'\i as (IGFAE) }
\email{javier.mas@usc.es}

\date{November $14^{th}$, 2021}

\definecolor{quantumviolet2}{HTML}{53257F}
\color{quantumviolet2}
\begin{document}
%\linenumbers
\begin{abstract}
An algorithm to classify a general Hermitian matrix according to its signature (positive semi-definite, negative or indefinite) is presented.  It builds on the  Quantum Phase Estimation algorithm, which stores the sign of the eigenvalues of a Hermitian matrix in one ancillary qubit. The signature of the matrix is extracted from the mean value of a spin operator in this single ancillary qubit. The algorithm is probabilistic, but it shows good performance, achieving 97\% of correct classifications with few qubits.  The computational cost  scales comparably to the classical one in the case of a generic matrix, but improves significantly for  restricted classes of matrices like $k$-local or sparse hamiltonians. 
\end{abstract}
\maketitle

\section{Introduction}\label{Section:Definiteness}

 The signature definiteness of a Hermitian operator is of mathematical and physical relevance. For instance, in optimization algorithms, examining the Hessian matrix permits to know if the current solution is a local minimum, a local maximum, or a saddle point \cite{Nocedal}. Also, positive semi-definiteness is one of the distinguishing features of a {\em bona fide} quantum density matrix.
 
 Classical methods to compute the definiteness of a Hermitian matrix $M$ are known in the literature. The Sylvester's criterion \cite{Meyer} asserts that $M$ is positive definite if {\em all  $N$  leading} principal minors are positive. However, to classify it as positive semi-definite, this property is not sufficient. In fact, {\em all} principal minors\footnote{The $k-$th leading principal minor $\Delta_k, k=1,...,N$ is obtained from $M$ by deleting the last $N-k$ rows and columns. Principal minors $D_k, 
~k=1,..., 2^N-1$, are instead obtained by deleting {\em any} $N-k$ rows and columns. } must be shown to be positive~\cite{Prussing1986}. This increases the cost considerably since $2
^N-1$ determinants have to be evaluated. Each one of them involves  $O(N!)$ operations, which can be reduced to $O(N^3)$ by, for example, Gaussian elimination. Another approach is Cholesky's decomposition \cite{Nocedal}, which also works only if $M$ is positive. Again, the computational complexity is $O(N^3)$\textcolor{red}, which, however, can be reduced to $O(N^{2.529})$ using the new matrix multiplication algorithms \cite{Camarero}. Finally, the obvious possibility is to calculate by brute force all the eigenvalues, which can be done in $O(N^w log_2(N))$ operations in some specific cases and $O(N^3 + (N\ log_2^2 N)\ log_2{b})$ in general~\cite{Pan1999}, being $2^{-b}$ the relative error. At present, the lowest theoretical bound is $w=2.376$. However, practical problems yield a higher lower bound $log_2(7)=2.808$.  This number will be the benchmark in our search for quantum advantage. 

In fact, despite being in its infancy, Quantum Computing has already provided valuable algorithms to deal with the eigenspectrum of operators \cite{Abrams1998,Shao2019,Dutkiewicz2021,Parrish2019,Gebhart2021,Wossnig_2018}. However, since calculating {\em all} the eigenvalues of a Hermitian NxN matrix is a complex problem \cite{Wocjan2006}, using these methods to compute its definiteness is not a sensible strategy to follow. A simpler problem was proposed in \cite{Somma2019} and coined as Quantum Eigenvalue Estimation Problem (QEEP). For a specific quantum state $\rho$ and  Hamiltonian $M$, it gives the probabilities of finding eigenvalues of $M$ supported by $\rho$ on each defined bin (of width $2\epsilon>0$, being $\epsilon$ the selected precision) of the discretized interval $[-1/2,1/2]$. Our proposal belongs to this class and focuses on a particular question. Namely, that of the character of the operator as described at the beginning of this section. In fact, the definiteness of a Hermitian matrix can be seen as a special QEEP, but for a state $\rho$ which is a linear combination of all the eigenvectors of the matrix $H$, and asking if the sum of the probabilities of all of the bins below 0 is 1, 0 or a value between. But this problem has two difficulties: we do not know $\rho$ in advance and, in general, the Hermitian matrices have eigenvalues $\lambda$ with $|\lambda|>1/2$. Finally, we want to mention the innovative scenario presented in \cite{martyn2021grand} where a particular application is advocated to give a solution to the so-called {\em eigenvalue threshold} problem, i.e., to know if the matrix has an eigenvalue lower than a defined threshold.  

The algorithm in this paper focuses only on the definiteness problem and is of a hybrid kind. It starts with a classical preprocessing that finds a higher (lower) bound to the highest (lowest) eigenvalue of $M$. Experimentally, in around $65\%$ of the cases, this stage will be enough to pin down  the definiteness of $M$ in $O(N^2)$ steps.  

% Because the needed initial state is unknown, Nevertheless, its outcome is peaked around the correct answer with an accuracy that increases with the depth $n$ of the ancillary system. For $n=14$ the final performance reaches $97\%$ of correct classifications. 

%To make a long story short, our solution 
For the remaining cases, our proposed quantum algorithm solves the classification problem, albeit in a non-deterministic way. It makes use of the Quantum Phase Estimation (QPE) algorithm \cite{nielsenchuang} where, in the end, only a single qubit of the ancillary system is inspected.  The measured value of $\langle \sigma_z\rangle$ in this particular qubit, determines if $M$ is positive semi-definite ({\sl iff} $\langle \sigma_z \rangle = 1$), negative definite ({\sl iff} $\langle \sigma_z \rangle = -1$), or indefinite for $\langle \sigma_z \rangle \in (-1,1)$. To be more precise: Let $n$ be the width of the ancillary system, the previous statements are exact {\sl iff} for all $i$, $2^n \lambda_i$ are integers. Otherwise, the value of $\langle\sigma_z\rangle$ will only reveal the correct definiteness in a probabilistic sense, whose accuracy can be enhanced arbitrarily by increasing $n$. %For the algorithm to work, the initial state should have overlap with all the eigenvectors of the investigated operator, $M$, much as in \cite{Abrams1998}. % 
%Using the previous classical results, the eigenvalues can be rescaled to be in the desired interval $[-1/2,1/2)$ where QPE works. To surpass the problem of the unknown initial state, we propose to sort this by running the circuit repeatedly with different initial states. Our simulations show that three initial vectors of the form $\ket{0}^{\otimes n}, \ket{1}^{\otimes n}$ and $\ket{+}^{\otimes n}$ are good enough to accomplish the task. Although probabilistic, its outcome is peaked around the correct answer with an accuracy that increases with the depth $n$ of the ancillary system. 
For $n=14$ the final performance reaches $97\%$ of correct classifications.

The paper is organized as follows. Section 2 summarizes the QPE algorithm for unitary operators, laying the playground for the sign estimation rules. Section 3 addresses the extension of the method to Hermitian matrices. In section 4 we combine the strength of the classical {\sl preprocessing} with the quantum sign estimation to yield a full hybrid algorithm whose strength is analyzed in section 5. There, the performance is presented as the result of numerical simulations. We wrap up with some comments and outlook in section 6.

%
% Sección Eigenvalue Estimation Algorithm
%

\section{Quantum Phase Sign Estimation}

In this section, we shall review the QPE algorithm with a slight twist that encompasses the core of the proposal.
Let $|u\rangle$ be an eigenstate of a unitary operator $U$. Unitarity implies that all the eigenvalues must be pure phases
\begin{equation}
U\ket{u} = e^{i 2\pi \theta } | u\rangle \, .
\label{eq:ph}
\end{equation}
Although in principle $\theta\in {\mathbb{R}}$ is a real number, periodicity of the phase implies  the equivalence relation $ \theta \sim  \theta + k, k\in {\mathbb{Z}}$. We can label each equivalence class by a unique representative $\theta \sim \tilde \theta$ lying in the fundamental domain $\tilde\theta\in [0,1)$.

\begin{figure}[ht]
\begin{center}
    \includegraphics[scale=1.1]{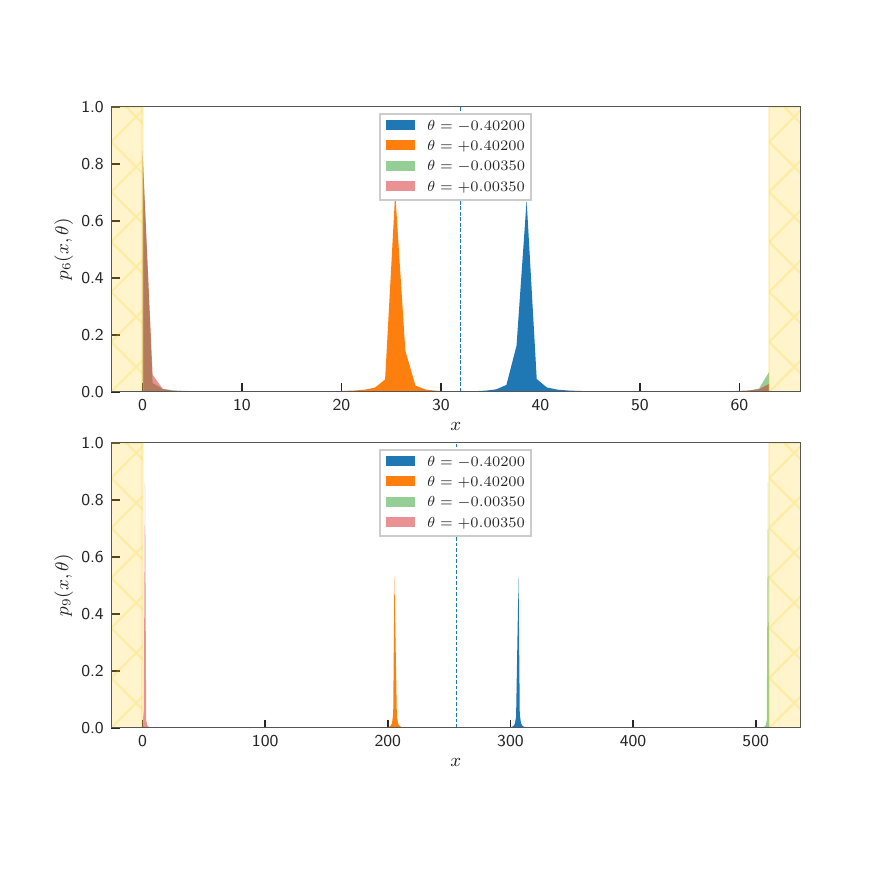}
    \caption{Probability function $p_n(x,\theta)$, of measuring the different states of the $n-$qubit ancilla register in the QPE algorithm for $n=6$ and $n=9$. The value $\theta = 0.00035$ is close to 0 and, hence, there is a substantial probability of producing a large error in $x$. By increasing $n$ from 6 to 9 the peaks sharpen enough to fit completely inside the same sign-subinterval as that of $\tilde\theta$. The brown color in the first plot comes from the overlap of the green and pink curves}
    \label{fig:Probability_theta}
\end{center}
\end{figure}

The QPE algorithm produces an approximation of order $n$ to  $\tilde\theta$ in the following sense: after initializing the circuit in the state 
$
     H|0\rangle^{\otimes n}\otimes|u\rangle
$
the output state, $\ket{\Phi}$, will have the ancillary part set in a superposition of the computational basis
$\ket{x}, x\in {0,...,2^{n}-1.}$
The probability of measuring a particular state $|x\rangle$ in the $n$-ancilla register peaks sharply around the integer part  $a=$Floor$[2^n \theta]$ as given by the following expression \cite{nielsenchuang}
\begin{equation}
   p_n(x,\theta) =\frac{1}{2^{2n}} \left\vert \frac{\sin \pi   (2^n\theta-x)}{\sin (\pi(2^n\theta-x)/2^n)} \right\vert^2 \, .  \label{eq:probdist}
\end{equation}
Figure \ref{fig:Probability_theta} illustrates the dependence of this probability distribution on the number of qubits of the ancilla register. Notice two distinctive features. On one hand, it  decays symmetrically at both sides of $a$, and the width shrinks rapidly with $n$. On the other, extending its domain to the full set of integers,  $x\in {\mathbb{Z}}$,  $p_n(x,\theta)$ has the {\em shift} periodicity $p_n(x,\theta) = p_n( x + 2^n k,\theta)$. Then, for any $\theta$, the value of $x$ is always in the canonical interval $x\in [0,2^n)$. It also has an important consequence for values of $\theta$, very close to $0$. For these cases, one side of the tail of the distribution may cross the limits. By the shift symmetry, this tail reappears at the opposite edge of the canonical interval, hence giving a significant probability to a {\em large} systematic error in the identification of $\theta$.
This is the key observation that underlies the proposal as we now want to discuss in detail.

Let us assume that some unknown phase variable, $\theta$, is said to take values in the interval $ [-0.5,0.5)$. To keep the discussion simple, we shall start by choosing a value of $\theta$ for which $2^n\theta \in {\mathbb{Z}}$ is an integer. In this case, the QPE algorithm provides an exact answer. If $\theta\in [0,0.5)$ is non-negative    then $\tilde \theta = \theta$ and the QPE algorithm will exactly retrieve the $n$-th order approximation $x = 2^n \tilde\theta \in [0, 2^{n-1})$. Notice that in such case, the most relevant qubit in $\ket{x}$ will be in state $\ket{0}_{n-1}$.\footnote{we label the $n$ qubits as $0,1,..,n-1$} If, however $\theta\in [-0.5,0)$ is negative,  QPE will produce an approximation for the equivalent representative $\tilde \theta = \theta + 1 \in [0.5,1)$. In this case $x = 2^n\tilde \theta \in [2^{n-1}, 2^n-1]$ will be such that the most relevant ancillary qubit in $\ket{x}$ will be in state $\ket{1}_{n-1}$. In summary, for values of $\theta \in [-0.5,0.5)$ such that $2^n\theta \in {\mathbb{Z}}$, the  eigenvalues $\sigma_z({n-1}) =\pm 1$ of $\sigma_z$ acting on the most significant ancillary qubit, correlate exactly with sign$(\theta)$. For obvious reasons, we will refer to the subintervals $[0,0.5)$ and $[0.5,1)$ where the representative $\tilde \theta$ can belong to, as positive and negative {\em sign-subintervals} respectively
\begin{equation}
\begin{array}{|cccccc|} \hline
 \rule{0mm}{5mm}  \theta \in [0,0.5)  & \stackrel{QPE}{\longrightarrow} & \tilde\theta \in [0,0.5) & \hbox{positive sign-subinterval} & \rightarrow & \sigma_z({n-1}) =+1 \\
\rule{0mm}{4mm}   \theta \in [-0.5,0)  & \stackrel{QPE}{\longrightarrow} & \tilde\theta \in [0.5,1) & \hbox{negative sign-subinterval} & \rightarrow & \sigma_z({n-1}) =-1
\\ 
\hline
\end{array}
\end{equation}
\iffalse
The apparent lack of generality in the choice of values for the original eigenvalue as restricted to lie in the interval $\theta \in [-0.5,0.5)$ will become clear below.
\fi

However, in general, $2^n\theta \notin {\mathbb{Z}}$. In this case, as commented before, the outcome $x$ follows a probability distribution which is sharply peaked around $a=\hbox{Floor}(2^n\theta)$. Concerning the possibility of retrieving sign$(\theta)$, there are two types of situations. The dangerous one occurs when $\theta$ is close to either $0$ or $\pm 0.5$. In any of these three cases, part of a side tail  in the probability distribution may cross these limits, giving chances to obtain values of $x$ in the wrong sign-subinterval, hence with the opposite sign to that of $\theta$. Turning the argument around, values of $\theta$ sufficiently detached from the boundaries, $0$ and $\pm 0.5$ will be such that both side tails of $p(x)$ will fit completely within the  sign-subinterval where $2^n\theta$ belongs. Let us stress that this pathology is inherited from the QPE protocol itself. In that case, only the values $\theta \sim 0, 1 $ were fragile in the sense of producing unreliable values of $x$. Now, for the sign, all four possibilities for $\theta \sim 0,\pm 0.5, 1$ are {\em fragile}, because the sign-subintervals that are retrieved for $x$ in those cases are prone to  misidentification.  

The QPE algorithm works under the assumption that we know an eigenstate of a unitary operator, $U$. It is still of use when we lift this hypothesis. Indeed,
if  $U$ is a  unitary $N\times N$ matrix with $N=2^m$, it will admit an orthonormal basis $\ket{u_i}, \,  i=0,...,N-1$ such that  $U = \hbox{diag}(e^{i2\pi \theta_0},...,e^{i2\pi \theta_{N-1}})$ for a set  of phases $\vec\theta = (\theta_0,...,\theta_{N-1})$.  Initializing the QPE circuit with a  generic vector    
$
\ket{b} = \sum_{i=0}^{N-1} \beta_i \ket{u_i}\, 
$
the probability of measuring a particular ket $\ket{x}$ at the end will  be
\begin{equation} 
p_n\left(x, \vec\theta\rule{0mm}{3mm}\right) =  \sum_{i=0}^{N-1} |\beta_i|^2 \, p_n(x,\theta_i) \, .
\label{eq:pxmi}
\end{equation}
This is nothing but a normalized sum over distributions of the form \eqref{eq:probdist} peaked at the different values of $2^n \theta_i,\, i=0,...,N-1$. Therefore, retrieving these phases $ \theta_i$ requires statistical sampling.

%\begin{figure}[h]
%    \centering
%    \includegraphics[scale=0.90]{Prob_vs_qubits.pdf}
%    \caption{Probability of measuring the sign of the eigenvalues versus the number of qubits. On the right, two eigenvalues that are large enough to have a good approximation with a low number of qubits because they are not close to zero. On the left, small eigenvalues which need a larger number of qubits to differentiate of zero. The probabilities change just when the number of qubits are just enough to separate them from zero.}
%    \label{fig:ProbVsQubits}
%\end{figure}

Regarding the signs, let us come back to the assumption that $\forall i$, $\theta_i \in [0,0.5) \,$ as well as $2^n \theta_i \in {\mathbb{Z}}$. Then it is clear that all the states $\ket{x}$ in the final  superposition
%\eqref{eq:phiensem} 
will have their most significant qubit in state $\ket{x}_{n-1}=\ket{0}_{n-1}$. Consistently, the expectation value of $\langle\sigma_z\rangle_{n-1} \equiv \langle \sigma_z(n-1)\rangle =  \bra{\Phi}\sigma_z\otimes {I}^{\otimes n-1}\ket{\Phi}$ will be equal to 1. 
The same conclusion can be reached for negative $\theta_i \in [-0.5,0)\, \forall i$, now yielding instead $\langle\sigma_z\rangle_{n-1}=-1$. Conversely, any value of $|\langle\sigma_z\rangle_{n-1}| < 1$ diagnoses the fact that the ensemble of phases $\theta_i$ has mixed signs. Summarizing, the expectation value $\langle\sigma_z\rangle_{n-1}$ allows to discriminate among three possibilities related to the sign distribution of the set of phases $\theta_i$.

When $2^n \theta_i \notin {\mathbb{Z}}$ has a nonvanishing mantissa, the above certainties turn into  probability distributions of finding either $\ket{0}$ or $\ket{1}$ when measuring the most significant ancillary qubit $\ket{x}_{n-1}$. Since all the states are orthonormal, this is obtained by simply  marginalizing overall values of $x$ compatible with $x_{n-1}=0$ or $1$.   
Turning these probabilities into expectation value of $\sigma_z$,  measured on the most significant qubit, we arrive at
\begin{equation}
\langle \sigma_z\rangle_{n-1} = 
\sum_{x=0}^{2^{n-1}-1} p_n\left(x, \vec\theta\rule{0mm}{3mm}\right)
-
\sum_{x=2^{n-1}}^{2^{n}-1} p_n\left(x, \vec\theta\rule{0mm}{3mm}\right). \label{eq:sigmazfinal}
\end{equation}
\begin{figure}[h]
    \centering
    \includegraphics[scale=0.8]{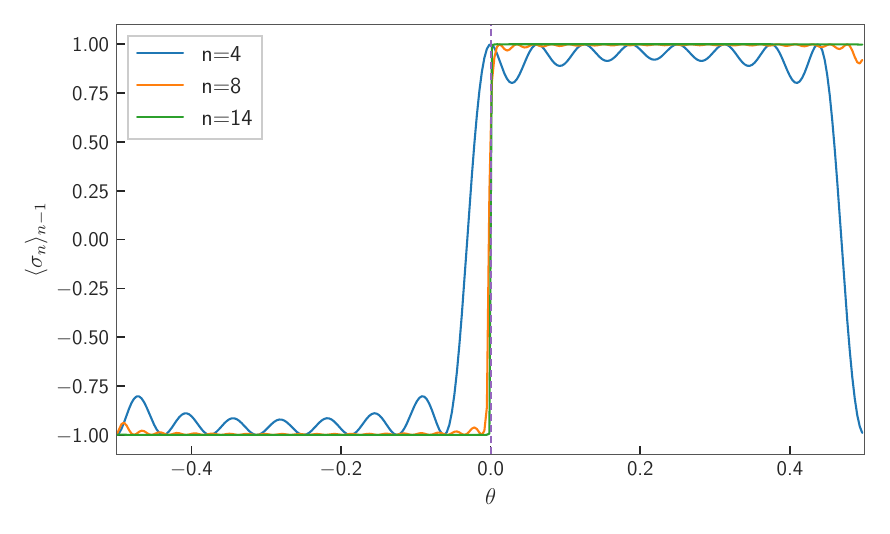}
    \caption{$\langle\sigma_z\rangle_{n-1}$ vs the value of a single eigenvalue for different number of qubits on the ancilla, following equation \eqref{eq:sigmazfinal}.  Notice that for $n=4$ there is a large value of $\langle\sigma_z\rangle_{n-1}$ for small negative values of $\theta$. This is avoided by increasing $n$. For $n\to \infty$ the curve tends to a step function jumping at 0.}
    \label{fig:SigmaZvsAngle}
\end{figure}
For the case of a single phase, $N=1$, \figref{fig:SigmaZvsAngle} shows the value of \eqref{eq:sigmazfinal} as a function of $\theta$ for different number of qubits in the ancilla, $n$. Increasing $n$  makes the curve tend to a perfect step function that discriminates sharply between negative and positive values of $\theta$. 

Before numerically analyzing this expression, let us finish the argument that is advertised in the title of this paper.

%%%%%%%%%%%%%%%%%%%%%%%%%%%%%%%%%%%%%%%%%%%%%%%%%%
\begin{figure}[t]
    \centering
    \resizebox{0.95\textwidth}{!}{%
    \begin{quantikz}
\lstick{$c_0 \ket{0}$} & \gate{H} & \ctrl{4} & \qw&\qw& \hfill\cdots\hfill& \qw & \gate[wires=4][2cm]{QFT^{\dagger}} &\qw\\
\lstick{$c_1 \ket{0}$} & \gate{H} &\qw& \ctrl{3} &\qw & \hfill\cdots\hfill&\qw& \qw & \qw \\
\lstick{\vdots}&&&&&& &&\\
\lstick{$c_{n-1}\ket{0}$} & \gate{H} & \qw& \qw & \qw&... & \ctrl{1} & \qw & \qw&\meter{} \rstick{$\sigma_z$} \\
\lstick{$b_0 \ket{0}$}&\gate[wires=3][2cm]{\text{Init to } \vec{b}}&\gate[wires=3][2cm]{e^{iMt_02^0}}&\gate[wires=3][2cm]{e^{iMt_02^1}}&\qw&\hfill\cdots\hfill&\gate[wires=3][2cm]{e^{iMt_02^{n-1}}} & \qw & \qw\\
\lstick{\vdots}&&&&&&\qw&\vdots&\\
\lstick{$b_{m-1} \ket{0}$}&\qw&\qw&...&\qw&\hfill\cdots\hfill&\qw & \qw & \qw 
\end{quantikz}
}%
\caption{Quantum circuit used to calculate the definiteness of a Hermitian matrix $M$}
    \label{fig:QPC}
\end{figure}
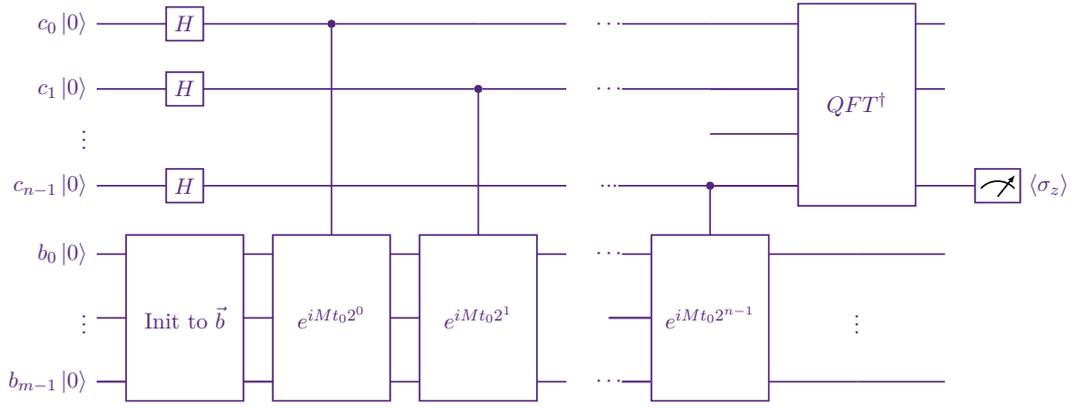
%%%%%%%%%%%%%%%%%%%%%%%%%%%%%%%%%%%%%%%%%%%%%%%%%%
%\begin{figure}
%    \centering
%    \includegraphics[scale=0.30]{circuito_ejemplo.pdf}
%\caption{Transpiled circuit example for 2 ancilla qubits using Qiskit. The \emph{Standard} version of the inverse of the Quantum Fourier Transform was used, with swapping, measuring $c_{n-1}$. If other inverses of QFT are used, the algorithm must take into account the final order of the qubits on the ancilla register.}
%    \label{fig:quiskit}
%\end{figure}

\section{Quantum Eigenvalue Sign Estimation for Hermitian matrices}\label{sec:QEE}

 A Hermitian matrix $M \in {\mathbb{C}}^{N \times N}$ has $N$ real eigenvalues $\lambda_i \in \mathbb{R}$, with  eigenvectors $\{|v_i\rangle, i=0,...,2^m-1\}$ which can be arranged to form a orthonormal basis.  Hereafter we will assume that $N=2^m$. Other cases can be treated as well by padding enough rows and columns with zeroes. 

The unitary evolution operator $U_M(t)=e^{iMt}$ has the same eigenvectors $|v_i\rangle$ as $M$,  with eigenvalues $\gamma_i = e^{i \lambda_i t }$ instead.
Hence the action of $U_M(t)$ on an arbitraty vector $|b\rangle = \sum_{i=0}^{2^m-1} \beta_i |v_i\rangle$ is 
\begin{equation} 
    U_M |b\rangle = \sum_{i=0}^{2^m-1} \beta_i U_M |v_i\rangle =  \sum_{i=0}^{2^m-1} \beta_i e^{it\lambda_i} |v_i\rangle \, .
    \label{eq:umdetm}
\end{equation}
The QPE algorithm maps values of $\theta_ i = \lambda_i t/2\pi$ onto states of an ancillary register. Now comes the key observation:  writing  $t=\frac{2\pi}{C}$, the constant $C$ can be adjusted so as to achieve  
\begin{equation}
\theta_i =\lambda_i/C\in [-0.5, 0.5)\, ,
\label{boundlambda}\end{equation}
for $i=0,...,2^m-1$.
This brings us back to the topic explored in the previous section, where the most significant ancillary qubit carries (with high probability) the information about the sign of $\theta_i$. 

Let us now discuss how to conveniently adjust the constant $C$, given the fact that the eigenvalues of $M$ are not known {\em a priori}.  All we need  is an upper bound on their absolute values that allows selecting $C$ such that \eqref{boundlambda} is fulfilled. In \cite{Wolkowicz1980} Wolkowicz and Styan derived bounds for the highest and lowest eigenvalues, $\lambda_{high}$ and  $\lambda_{low}$ respectively, of an arbitrary  matrix only in terms of its trace. Let us define the quantities
\begin{equation}\label{eq:m}
r=\frac{Tr(M)}{N}~~~~, ~~~~s= \sqrt{\frac{Tr(M^2)}{N}- r^2} \, .
\end{equation}
The following bounds are proven for $\lambda_{low}$ 
\begin{equation}
C^{min}_{low}\leq \lambda_{low} \leq C^{max}_{low} ~~\, , ~~~ 
\label{eq:llow}
\end{equation}
with 
\begin{equation} 
C^{min}_{low}= r-s\sqrt{N-1}~~~,~~~C^{max}_{low} = r - \frac{s}{\sqrt{N-1}} \, ,
\label{eq:Cllow}\end{equation}
and similarly for $\lambda_{high}$
\begin{equation}
C^{min}_{high}\leq \lambda_{high} \leq C^{max}_{high} ~~~\, , ~~~
\label{eq:lhigh}
\end{equation}
with 
\begin{equation}
C^{min}_{high}= r +\frac{s}{\sqrt{N-1}} ~~~ ,~~~ C^{max}_{high} = r+s\sqrt{N-1} \, .
\label{eq:Clhigh}
\end{equation}
Hence all we need to do is compute some traces. The complexity is dominated by the calculation of $Tr(M^2)$ which needs $O(N^2)$ operations. 
\iffalse
for $M$ a \textcolor{red}{Hermitian matrix}:

\begin{equation}\label{eq:trace}
Tr(M^2) = \sum_{i,j} (M \odot M^T)_{ij}\, ,
\end{equation}
\todo[color=\Javier]{yo sigo sin ver este "truco"} 
which reduces to $O(N^2)$ the complexity of this operation. 
\fi
Now clearly, choosing  
\begin{equation} 
   C =  2 \,{\rm Max}(\, |C^{min}_{low}|\, ,\,  |C^{max}_{high}| \, ) \, ,
   \label{eq:Cbound}
\end{equation}
all the phases will satisfy 
\begin{equation}
|\theta_i| = \left\vert\frac{\lambda_i}{C}\right\vert \leq 0.5 \, .
\end{equation}
Summarizing: in what concerns our original problem, non-negative  eigenvalues $\lambda_i\geq  0$ have been mapped onto $\theta_i \in [0,0.5)$ and QPE will retrieve a value $\tilde \theta$ in the same positive sign subinterval. On the other hand, negative eigenvalues $\lambda<0$ will be mapped onto $\theta\in [-0.5,0)$. QPE will give a representative $\tilde\theta\in [0.5,1)$ on the negative sign-subinterval. Measuring $\langle \sigma_z\rangle_{n-1}$ will reveal this sign-subinterval to a given accuracy.

\section{Hybrid algorithm for classification of Hermitian matrices}

Putting together all the previous results, we can formulate the hybrid algorithm as a two-stage protocol. Given a Hermitian matrix $M$, to classify it according to its definiteness:
\begin{enumerate}    
    \item perform a classical check making use of the bounds in eqs. \eqref{eq:llow} and \eqref{eq:lhigh}.  The subroutine that performs this task is shown in Algorithm \ref{algo:Classical}. If the classical test succeeds, finish. Otherwise, namely if the classical test fails, it is necessary to proceed with step 2.
    \item Because $M$ could not be classified classically, proceed with the quantum part, which will anyway need the bounds computed in step 1, and which is summarized in Algorithm \ref{algo:quantum}. 
\end{enumerate}

%Our results indicate that, for a random sample of hermitian matrices $M$, around $60-70 \%$ could be classified by the classical algorithm with full accuracy. From the rest, a $95\%$ was satisfactorily classified by the optimally tuned quantum circuit. Altogether, the accuracy of the hybrid algorithm reaches $97\%$.

The classical step  makes use of equations \eqref{eq:llow} to \eqref{eq:Clhigh}. For example, if $C^{min}_{low}> 0$  then the minimum eigenvalue will be positive and the matrix $M$, accordingly,  positive definite. The full list of possibilities is evaluated in Algorithm \ref{algo:Classical}  as follows
\begin{equation}
\begin{array}{rcl}
  C^{min}_{low}   & \longrightarrow &
  \left\{ 
 \begin{array}{ccc} 
 >0 &\hbox{positive definite}\\
 =0 & \hbox{positive semi-definite}\\
 \end{array}\right.
\\
  C^{max}_{high}   & \longrightarrow &
  \left\{ 
 \begin{array}{cc} 
 <0 &\hbox{negative definite}\\
 =0 &\hbox{negative semi-definite}\\
 \end{array}\right.
 \\
 \rule{0mm}{6mm}
  C^{max}_{low}<0 ~~  \hbox{and} ~~  C^{min}_{high}>0    & \longrightarrow &  \hbox{indefinite}\, .\\
 \end{array} \label{eq:shortcut}
\end{equation}

 Outside these situations, the character of $M$ cannot be identified without ambiguity and the quantum algorithm described in the previous section must be used.
As explained, the quantum approach proceeds in two steps. From the initial matrix $M$, whose eigenvalues are $\lambda_i,\,  i=0,...,2^m-1$, by a suitable rescaling we can obtain another one $\tilde M = M/C$ whose eigenvalues belong to the desired range $\theta_ i = \lambda_i/C \in [-0.5,0.5)$. The definiteness of $\tilde M$ and $M$ are the same. However, for $\tilde M$, its definiteness can be inferred from the measurement of $\langle \sigma_z \rangle_{n-1}$.  Let us expand a little bit on the details of both steps.

In what concerns the first part, namely the rescaling, all we need to know is  the constant $C$. Here is where the work done before  for the classical stage is reusable and $C$ will be given  by the expression \eqref{eq:Cbound}.
At a practical level, as mentioned in eq. \eqref{eq:umdetm} this constant is included in the algorithm in the selection of the Hamiltonian evolution time step
\begin{equation}
    U(t_0) = \exp(i t_0 M) 
    \label{unitaryM}
\end{equation}
by setting  $t_0 = 2\pi/C.$ 

%
% Algoritmo clasico
%
\begin{algorithm}
\SetKwFunction{FRecurs}{Function \texttt{checkClassical}}
\SetFuncSty{textbf}

\SetAlgoVlined
%\LinesNumbered

\begin{tabular}{ll}
\KwIn{}&{$\boldsymbol{M} \in \mathbf{C}^{N\times N}$ Hermitian Matrix }\\
\KwOut{}&Flag with class or error.\\
&Low and high limits for the  Eigenvalues of the Matrix $\boldsymbol{M}$
\end{tabular}
\BlankLine

\If{$M \ne M^{\dagger} \hbox{(up to atol)}$}{\Return{\text{Error}}}
\tcp{Initialize internal variables}
{$N \leftarrow $dimension$(M)$}\\
{$r  \leftarrow \text{using \eqref{eq:m}}$}\\
{$t \leftarrow Tr(M^2)$}\\
{$s \leftarrow \text{using \eqref{eq:m} and } t$}\\
 {$limitLowMin \leftarrow r-s\sqrt{N-1}$}\\
 {$limitLowMax \leftarrow r-\frac{s}{\sqrt{N-1}}$}\\
 {$limitHighMin \leftarrow r+\frac{s}{\sqrt{N-1}}$}\\
 {$limitHighMax \leftarrow r+s\sqrt{N-1}$}\\
\tcp{Classify}

\uIf{limitHighMax < 0}
    {$classM \leftarrow$ \text{Flag Negative Definite}}
\Else{
      \uIf{limitHighMax $\le$ 0}
                    {$classM \leftarrow$ \text{Flag Negative Semi-Definite}}
      \Else{
            \uIf{limitLowMin $>$ 0}
                    {$classM \leftarrow$ \text{Flag Positive Definite}}
            \Else{
                    \uIf{limitLowMin $\ge$ 0}
                            {$classM \leftarrow$ \text{Flag Positive Semi-Definite}}
                    \Else{
                            \uIf{(limitHighMin $>$ 0) $\&$ (limitLowMax $<$ 0)}
                                    {$classM \leftarrow$ \text{Flag Indefinite}}
                            \Else{$classM \leftarrow$ \text{Flag Unclassified}}
                        }
                }
            }
     }
\Return{classM,  limitLowMin, limitHighMax}
\caption{Classical check of definiteness of a Hermitian matrix $M$ using its trace.}\label{algo:Classical}
\end{algorithm}
\begin{algorithm}
\SetKwFunction{FRecurs}{Function \texttt{checkQuantum}}
\SetFuncSty{textbf}

\SetKwComment{Blank}{***** }{*****}

\SetAlgoVlined
\LinesNumbered

\begin{tabular}{ll}
\KwIn{}&$\boldsymbol{M} \in \mathbf{C}^{N\times N}$ Hermitian Matrix\\
&$\boldsymbol{limitLowMin}$, the lower limit for the eigenvalues\\
&$\boldsymbol{limitHighMax}$, the higher limit for the Eigenvalues\\
&$\boldsymbol{Boundary}$, the boundary for the classification ($\delta$)\\
&$\boldsymbol{Trials}$, number of trials\\
&$\boldsymbol{n}$, number of qubits for the ancilla register in the QPE algorithm\\
&$\boldsymbol{atol}$, tolerance to distinguish one number from 0\\
&$\boldsymbol{Shots}$, number of shots for each trial\\

\KwOut{}&Flag with class of $\boldsymbol{M}$ or $\boldsymbol{Error}$
\end{tabular}
\BlankLine
\If{$M \ne M^{\dagger}$ }{\Return{\text{Error}}}
\If{(limitLowMin = None) or (limitHighMax = None)}{limitLowMin, limitHighMax $\leftarrow$ using \eqref{eq:llow} and \eqref{eq:lhigh}}
\If{{\rm dimension}($M$) {\rm mod}\,2 $\ne$ 0}{$d \leftarrow $ dimension($M$)\\expand $M$ with zeros so dimension($M$) mod 2 = 0}
{$N \leftarrow  $dimension$(M)$}\\
{$C \leftarrow  2\hspace{4pt} {\rm max}(|limitHighMax|,|limitLowMin|)$}\\
{$time \leftarrow \frac{2\pi}{C}$}\\
{$sigma \leftarrow 0$}\\
\For{$i\leftarrow 1$ \KwTo $Trials$}{
    $\overrightarrow{b} \leftarrow $normalized random vector $ \in {C}^N$ with N-d zeros at end\\
    Allocate ancilla register with $n$ qubits\\
    Allocate $|b\rangle$ register with $\log(N)$ qubits\\
    Initialize $|b\rangle$ to $\overrightarrow{b}$\\
    QPE on $M$ for $-time$ using time evolution. (Figure \ref{fig:QPC})\\
    $sigma \leftarrow sigma + \langle\sigma_z^{(n-1)}\rangle$ for Shots\\
}
{$classM \leftarrow$ \text{Flag Indefinite}}\\
\If{(sigma/Trials) $>=$ Boundary}{$classM \leftarrow \text{Flag Positive Semi-Definite}$}
\If{(sigma/Trials) $<=$ -Boundary}{$classM \leftarrow \text{Flag Negative Definite}$}

\Return{ classM}

\caption{Quantum classification of the definiteness of a Hermitian matrix}\label{algo:quantum}
\end{algorithm}

%
% Description quantum algorithm
%

In what concerns the computation of $\langle \sigma_z \rangle_{n-1}$, its reliability hinges upon two factors. The first one is the accuracy, which is related to the number of qubits, $n$, in the ancillary system. Increasing $n$ squeezes the probability distribution for the outcome $x$, as explained in Fig. \ref{fig:Probability_theta}, in such a way that it  fits almost completely within the same sign-subinterval as that associated with the original eigenvalue. The second one is the  state $|b\rangle$. In order to tag all the eigenvalues, it should have nontrivial projection onto all the basis eigenvectors $|v_i\rangle$.  Using a single random initialization of $|b\rangle$, there is a high probability that the decomposition includes all of them, but this is not assured. In fact, we have seen that, with a single initialization, the possibility of wrong classification is high. So, the strategy is to make several trials with different initial random vectors and calculate the average of the results for decision making. 

Finally, the classification  will depend upon a  threshold  which we will calibrate phenomenologically for optimal retrieval. Introducing a {\em cut} number, $\delta \in [0,1]$,  the quantum algorithm boils down to the following protocol: 
\begin{itemize}
    \item Run a number of times the circuit drawn in \figref{fig:QPC} which parametrically depends on $M$ and $t_0=2\pi/C$,  measuring always  $\overline{\langle\sigma_z\rangle}$ on the most relevant $(n-1)$-th ancillary qubit.
    \item Classify $M$ according to the following table
\begin{equation}
\overline{\langle \sigma_z \rangle}_{n-1} ~\in~
\left\{
\begin{array}{rcl}
[\delta,1] & \rightarrow & \hbox{positive semi-definite}\\
\rule{0mm}{4mm} [-1,-\delta] & \rightarrow & \hbox{negative definite}\\
\rule{0mm}{4mm} (-\delta,+\delta) & \rightarrow & \hbox{ indefinite}\\\end{array}
\right. \, 
\end{equation}

\end{itemize}
Ideally, we would like to send $\delta \to 1$ but this requires that $n\to \infty$ of the ancillary system.  The optimal value of $\delta$  will need to be fixed experimentally, and we will accomplish this task in the following section.

A final word regarding the case of Hermitian matrices whose dimension is not a power of 2, so that they cannot be represented within a qubit circuit.   In this case, the input matrix can be appended with vanishing lines and rows until a dimension $N=2^m$ is reached.  The random vector $\ket{b}$ will also include additional zeros in the last elements to match the correct dimension, while having vanishing projection on the appended null eigenspace.

%
% Results
%

\section{Results}

We start by summarising the results for readers who prefer to skip the details of the statistical analysis below.  The optimal value of $\delta$ is found to be $0.98$, which allows for a $\geq \! 97\%$ efficiency in the final retrieval of the correct definiteness, after averaging over at least three independent input states $\ket{b}$.

To derive this result from the algorithm, a sample of 1800 4x4 Hermitian matrices $M$ was generated. They were distributed equally over the three classes, with their eigenvalues chosen randomly and uniformly in the interval $[-1,1)$. 
In the positive semi-definite class, 5\% of the matrices were forced to have one eigenvalue equal to 0.

\begin{figure}[ht]
    \centering
    \includegraphics[scale=0.95]{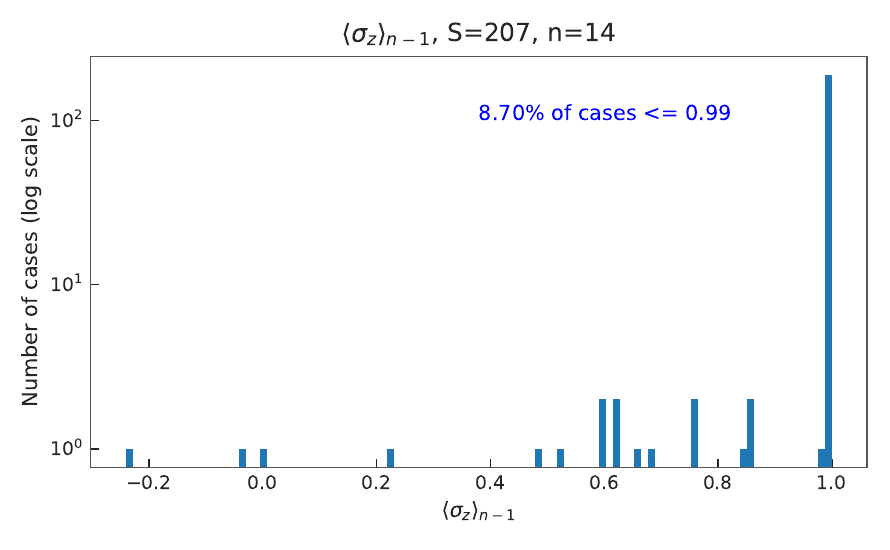}
    \caption{Results of 100 measurements for 207 positive definite matrices}
    \label{fig:sigma}
\end{figure}

For each matrix, the quantum algorithm was executed taking 100 as the number of shots to compute the average value of $\overline{\langle \sigma_z\rangle}_{n-1}$, and 5 trials for $\ket{b}$, saving the results of the trials for post-execution analysis. The algorithm was executed with $n=$ 4, 6, 8, 10, 12, and 14 qubits on the ancilla register. It was simulated without noise using Qiskit version 0.16.1~\cite{Qiskit} and BasicAer \emph{qasm\_simulator} as the provider. To simulate the time evolution of the Quantum Phase Estimation algorithm, the function \emph{evolution\_instruction}  that implements $U(t_0)^{2^k} = (e^{-i t_0 M})^{ 2^k}$ was used with one time slice  on the Trotter-Suzuki decomposition per $U(t_0)$. This is equivalent to simulating the time evolution of the operator $U(t_0 2
^k) = e^{-i t_0 M 2^k}$ with $2^k$ time slices on the Trotter-Suzuki decomposition.\footnote{The simulations were executed on CESGA FinisTerrae supercomputer between February and August of 2020, using Python 3.8.1 and Numpy 1.18.1.}

\begin{figure}[ht]
    \centering
    \includegraphics[scale=0.99]{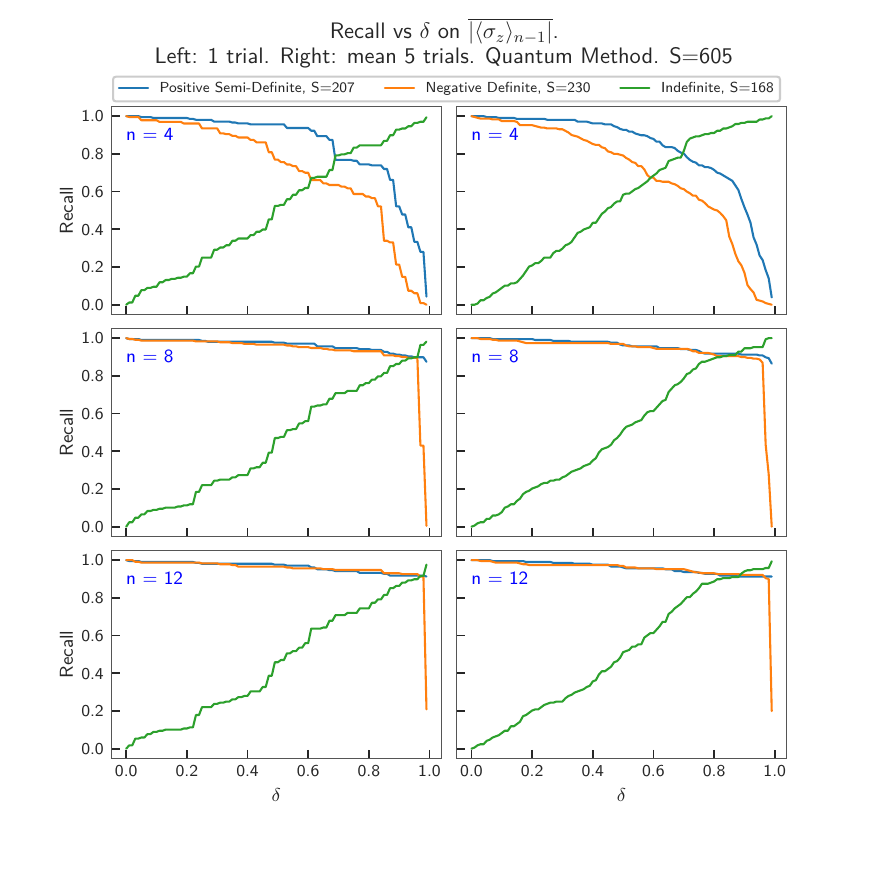}
    \caption{{\em Recall} of classification for each class for one and five trials (left and right resp.). Averaging of over trials enhances the {\em recall}  for the class of indefinite matrices. S is the number of  matrices of the sample.}
    \label{fig:fivetrials}
\end{figure}

In \figref{fig:sigma}, the results for a single initialization of $\ket{b}$ are shown in the case where $M$ is a positive definite matrix. Here $\langle\sigma_z\rangle_{n-1}=1$ was found in $91.3\%$ of the cases. There is still an 8.7\% of cases with values of this expectation value smaller than the maximum, meaning that some  $|1\rangle_{n-1}$ states have been measured, yielding an erroneous classification. Lowering the cut, $\delta <1 $, would increase the number of correctly classified matrices.

Our task is to adjust $\delta$ to maximize the classification capability over all the classes by using the metric {\em recall} = true positives/(true positives + false negatives). 
In \figref{fig:fivetrials} we plot the {\em recall} for the three classes as a function of the cut $\delta$.  We notice that, as $\delta$ is lowered from its maximum value, the value of recall  increases  for the  positive semi-definite and negative definite classes. However, it  decreases for the class of indefinite matrices, hence a compromise value has to be reached. On the left side plots, we show the results of a single initialization $\ket{b}$ for $n=4,8$ and $12$ qubits. A higher number of trials, $\{|b_i\rangle \}$, makes the results improve as shown in the column on the right. For $n\geq 12$ and 5 trials, the {\em recall} on all the classes comes very close to the maximum value in a region for $\delta \in [0,95,0.98]$. Our choice for $\delta = 0.98$ maximizes the  {\em accuracy} defined as the number of properly classified matrices divided by the total number of matrices in the sample, reaching 92.89\%.  This number is roughly the average value of {\em recall} on the three classes, whose values can be seen in  Table \ref{tab:accuracy_quantum}.

%%\begin{figure}
%%    \centering
%%    \includegraphics{Indefinite_vs_qubits.pdf}
%%    \caption{Dependence of the classification of indefinite matrices with the number of qubits and number of trials}
%%    \label{fig:Indefinite}
%%\end{figure}

\begin{figure}[ht]
    \centering
    \includegraphics[scale=0.99]{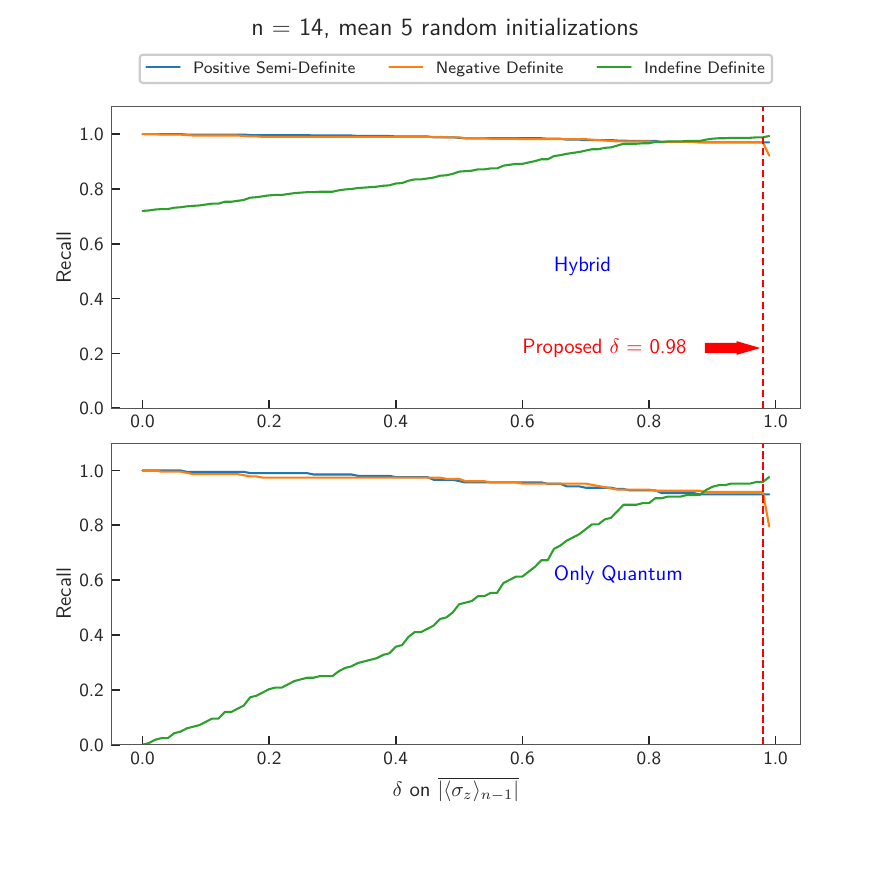}
    \caption{Classification boundary for 5 trials. Top, {\em recall} for classical-quantum hybrid algorithm. Bottom, {\em recall} for only quantum version.}
    \label{fig:Indefinite14qubits}
\end{figure}

A final improvement comes from taking advantage of the hybrid algorithm.  On the sample of 1800 4x4 matrices, the classical part could already correctly classify 64.67\% of the positive semi-definite matrices, 64.50\% of the negative cases, and 71.16\% of the indefinite ones. The modified curves for {\em recall} as a function of $\delta$ on the three classes for the combined hybrid algorithm  can be observed in the  \figref{fig:Indefinite14qubits} for the particular case of $n=14$ ancillary qubits. To appreciate the improvement  we have added the only-quantum version below.   Table~\ref{tab:accuracy_all} summarizes the new values of the {\em recalls} and global accuracy for the selected value $\delta =0.98$ again for $n=14$ for comparison with  Table \ref{tab:accuracy_quantum}. The global  increase in accuracy, up to  97.61\% illustrates the advantage of the  classical-quantum hybrid algorithm.

\begin{figure}
    \centering
    \includegraphics[scale=0.99]{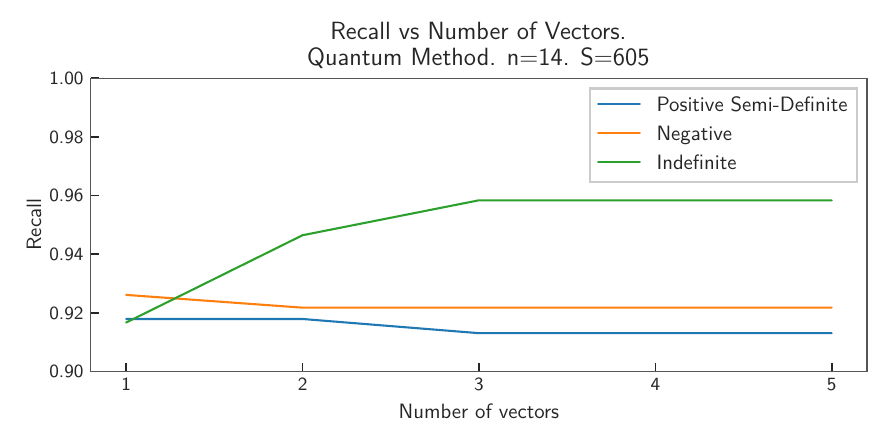}
    \caption{Behavior of {\em recall} vs. the number of initialization trials $\ket{b}$. We observe that  3 seems to be the optimal value and going beyond does not bring in any improvement. S is the number of matrices of the sample.}
    \label{fig:initb}
\end{figure}

Finally, the importance of initializing with more than one vector $\ket{b}$ is shown in \figref{fig:initb}. As a matter of fact, this plot reveals that going beyond 3 independent vectors yields no significant improvement. We then checked that initializing with  $\ket{0}^{\otimes n},\ket{1}^{\otimes n}$ and $\ket{+}^{\otimes n}$ performs as good as with three random vectors.

\begin{table}
    \centering
    \begin{tabular}{c|c|c|c|c}
    \hline
\textbf{Qubits ancilla}& \textbf{Recall}&\textbf{Recall}&\textbf{Recall }&\textbf{Accuracy(\%)}\\
\textbf{}& \textbf{Positive Semi-definite}&\textbf{Negative}&\textbf{ Indefinite}&\textbf{}\\
\hline4&0.14&0.00&0.99&32.40\\
6&0.83&0.01&1.00&56.36\\
8&0.89&0.28&1.00&68.93\\
10&0.91&0.83&0.98&89.59\\
12&0.91&0.90&0.96&92.07\\
14&0.91&0.92&0.96&92.89
    \end{tabular}
    \caption{Quantum algorithm: {\em recall} for each class and accuracy (\%) of the quantum classification versus the number of qubits on ancilla register for 5 trials and optimized cut  $\delta=0.98$.}
    \label{tab:accuracy_quantum}
\end{table}
\begin{table}
    \centering
    \begin{tabular}{c|c|c|c|c}
    \hline
\textbf{Qubits ancilla}& \textbf{Recall}&\textbf{Recall}&\textbf{Recall }&\textbf{Accuracy(\%)}\\
\textbf{}& \textbf{Positive Semi-definite}&\textbf{Negative}&\textbf{ Indefinite}&\textbf{}\\
\hline4&0.70&0.62&1.00&77.28\\
6&0.94&0.62&1.00&85.33\\
8&0.96&0.72&1.00&89.56\\
10&0.97&0.93&0.99&96.50\\
12&0.97&0.96&0.99&97.33\\
14&0.97&0.97&0.99&97.61
    \end{tabular}
    \caption{Full hybrid algorithm: {\em recall} for each class and accuracy (\%) of the classification versus the number of qubits on ancilla register for 5 trials and optimized value of  to $\delta=0.98$.}
    \label{tab:accuracy_all}
\end{table}

%\section{Experimental Results}

%Now let us address the full hybrid algorithm. As mentioned elsewhere above, the classical criteria was only capable of classifying correctly 65.50\% of the Positive Semi-Definite matrices, 61.67\% of the Negative cases and 72.00\% of the Indefinite ones. This gives an accuracy of 66.39\% if the classical part of the hybrid algorithm is only used. 
%On its side, the accuracy of the quantum algorithm strongly depends on the number of ancillary qubits (see last column in Table~\ref{tab:accuracy_quantum}). The interesting result is that they seem to be efficient in complementary subsets of the sample. Indeed, the final accuracy  is slightly better if the hybrid two-step algorithm is used as compared with either of them separately. This is reflected in  Table~\ref{tab:accuracy_all}. For example, 
%using 8 qubits on the ancilla register, an accuracy of 97.06\% is achieved with the hybrid algorithm, decreasing to 91.24\% when only the quantum part is used. These numbers demonstrate the advantage of adding the fast classical part to the quantum algorithm, to generate a more robust hybrid algorithm.

%
%
% Discusion
%
\section{Discussion and further work}
We have described a hybrid algorithm to classify Hermitian matrices according to their definiteness. 
It is probabilistic, although the achieved accuracy easily reaches 97\% for 4x4 matrices. A shortcoming seems to be the inability to distinguish between strictly positive and positive semi-definite cases. This limitation can be overcome by running the algorithm a second time, now for the negative matrix $M'=-M$. If the $M'$ is classified as negative definite (indefinite), the original matrix $M$ is positive definite (positive semi-definite).  

An analysis of the computational cost of the algorithm is pertinent. As said in the introduction,  for a Hermitian matrix of dimension $N$, the computational cost to calculate all the eigenvalues is  $O(N^w log_2(N))$, with $w= 2.376$ in the best case. However, usually, the classical methods have a complexity higher than $O(N^3)$. 
In our case, there is a minimum cost of $O(N^2)$ that  comes from the classical evaluation of the bounds \eqref{eq:llow} on the spectrum.
For 4x4 matrices, this classical stage correctly classifies $M$ in $\sim 65\%$ of cases, so this step sets the absolute lower bound for scaling for a generic matrix.

 Concerning the quantum algorithm, we need to independently evaluate the three intervening steps: the initialization of the state $|b\rangle$, the unitary time evolutions, and the QFT, both implicit in the QPE protocol. 
\begin{figure}[ht]
    \centering
    \includegraphics[scale=0.99]{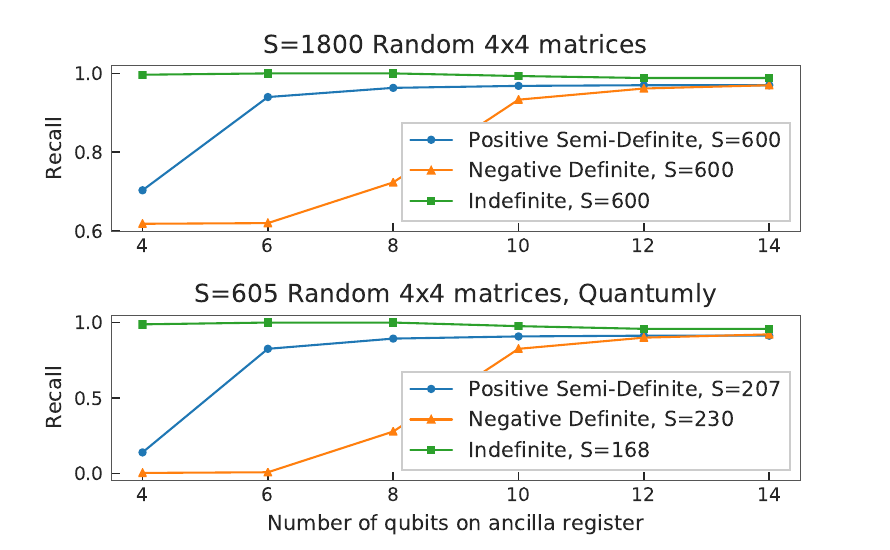}
    \caption{{\em Recall} of the classification for each class versus the number of ancillary qubits for random Hermitian matrices $\in {\mathbb{C}}^{4\times4}$. Top, the results of the classical-quantum hybrid algorithm. Bottom, the results of the fraction of the matrices which were classified using the quantum algorithm. Beware the ranges of {\em recall} in each plot. S is the number of matrices of each sample.}
    \label{fig:Accuracyvsqubits}
\end{figure}

The initialization of the random state $|b\rangle$ in the case of matrices with dimensions $N=2^m$ can be done using $ m = O(log(N))$ operations and  similar order  when the dimension of the matrix is not a power of two. The computational cost of the inverse QFT is $O(n^2)$, where $n$ is the number of qubits of the ancilla register, and is independent of the dimension of the matrix.\footnote{Actually, for an approximate Fourier transform as needed in the present context, this estimation is closer to $O(n\log n)$ as shown in \cite{nam2020}.} 

As for the $n$ unitary evolutions $U(t_0 2^k) = e^{-i t_0 M 2^k}, k=0,...,n-1$, the Trotter-Suzuky decomposition  has been chosen to be in $2^k$ uniform time slices. Refining this slicing must be studied with care because the eventual improvement in performance comes at the cost of a significant increase in  execution time. 
The main contribution to the scaling of the computational complexity comes from implementing these time evolutions. Nowadays, algorithms to compute it for a Hermitian matrix start by decomposing it into combinations of tensor products of Pauli matrices. For a matrix of dimension $N$, there are $4^{log_2(N)}=N^2$ possible terms in  this decomposition, obtained by calculating the trace of the product of each Pauli combination with the matrix. Because any possible tensor product of Pauli's operators has $N$ non-zero elements, the number of operations to prepare the matrix is $O(N4^{log_2(N)})=O(N^3)$. To implement this decomposition on a computable circuit, a maximum of $2(log_2(N)-2)$ CNOT gates and another $(2log_2(N)-1)$ rotations are needed~\cite{Gui}. This circuit must be executed $O(2^n)$ times. Putting all together, the computational complexity is $O(N4^{log_2(N)} + 2^{n+2}log_2(N)4^{log_2(N)})$, that it is dominated by the first term. In other words, on the current state of the technology, for a general Hermitian matrix, the computational complexity  grows as $O(N^3)$ for a fixed ancillary system.

\iffalse
Let us  now speculate about the  theoretical limits to the complexity of the quantum algorithm.  Certainly the bulk of it stems from the computational cost of  implementing the unitary $U$. Some results in the literature that hint towards the possibility of improving the scalings expressed above. For the case of a Universal Quantum Computer, Shende et al.~\cite{Shende2006} demonstrated that any unitary operator can be implemented on a quantum circuit containing no more than $O(N^2)$ CNOT gates.
More specifically, referring to the case of a Trapped Ion Computer,  Goubault De Brugi{\`{e}}re et al. \cite{GoubaultDeBrugiere2020} have shown that almost any unitary operation requires a number of gates \footnote{In Trapped Ion Computers the universal set of gates is composed of local $R_z(\varphi)$ and $R_x(\theta)$ gates, an the entangling M√∏lmer‚ÄìS√∏rensen gate defined by
$
    MS(\theta) = e^{-i\theta (\sum_{i=0}^{m}\sigma_x^i)^2/4}
$. } that is bounded by
%\begin{equation}
%    \#MS\ge \Biggl\lceil{\frac{4^m-3m-1}{2m+1}}\Biggr\rceil
%\label{eq:upper}
%\end{equation}
%being m the number of qubits and MS 
%Using these gates plus $R_x$ and $R_z$, almost any unitary operation can be implemented on a number of gates with a lower bounds given by:
\begin{equation}
    \#Gates\ge (4m+1)\Biggl\lceil{\frac{4^m-3m-1}{2m+1}}\Biggr\rceil + 5m \approx O(N^2) \, .
\label{eq:upper2}
\end{equation}
These results point out towards the possibility of finding a bound for the computational cost of the quantum part of our algorithm  below the naive counting ${\cal O}(N^3)$. 
\fi

As a consequence, the proposed quantum algorithm has a computational complexity scaling that improves over that of the actual classical algorithms, but is worse than their classical theoretical limit. Notice, however, that we have analyzed the worst possible situation given by a generic structureless Hermitian matrix $M$. 
 Already the initialization implies order $N^2$ operations, as high a number as for  decomposing $M$ in Pauli strings, required by the time evolution. However, when applied to restricted classes of matrices of frequent use, like sparse matrices or k-local hamiltonians, the complexity cost will drop dramatically. In some instances, $M$ will already be given by a sum of Pauli strings, or be an Ising like hamiltonian, in which case, even the classical estimation of the bounds on the spectrum could \iffalse can\fi be done in \iffalse much\fi less than $N^2$ steps. In these cases, the computational complexity of the quantum algorithm will be dominated by the QPE and not by the classical pre-processing, thereby decreasing to $O(2^{n+2}log_2(N)4^{log_2(N)})$.  Anyway, we will not insist further on these subtleties as long as the algorithm presented here is generic, and treats all cases in the same way. 
 
 Because QPE is not NISQ-friendly, due to its dependence on the QFT and on the time evolution, an open question is if new algorithms for phase estimation \cite{Dutkiewicz2021,Clinton2021,Kivlichan2019,Wan2021,OBrien2019,martyn2021grand} or improved time evolution methods (using variational algorithms \cite{Cirstoiu2020,Commeau2020}, truncated Taylor series \cite{Berry2015}, linear combination of unitaries \cite{Childs2012}, or qubitization \cite{Low2019}) can be used to solve the definiteness problem. In some of these cases, when block encoding of the Hermitian matrix is required, equation \eqref{eq:Cbound} could be useful to permit such encoding. These possibilities are open for further research.

Summarizing: the algorithm  we present here, is composed of a classical part and a quantum part whose combined performance reaches easily 97\% with a computational cost that, within the current state of the technology, is no worse than that of the purely classical one, leaving room for improvement with the foreseeable  optimization in two key steps, the initialization and  the implementation of the time evolution operator.

\section{Data Availability}
The datasets generated and analysed during the current study are available from the corresponding author on reasonable request.

%
% Agradecimientos
%
\section{Acknowledgements}
The authors want to thank CESGA for  provisioning of the computing resources, specially for the computing time on FinisTerrae II. Also, the authors have used extensively the computing resources integrated in this supercomputer from Instituto Gallego de Física de Altas Energías (IGFAE). Those resources have been funded by Axencia Galega de Innovación (GAIN). Without them, this work would not have been possible. We also appreciate the comments from Cyril Allouche and Timothee Goubault de Brugiere from ATOS, which help to improve the content  significantly, from Juan José Nieto Roig from the University of Santiago de Compostela, from Elias F. Combarro from the University of Oviedo, and from Alejandro Gómez Frieiro from the University of Queensland. 
The work of JM was supported by grants FPA2014-52218-P  and FPA2017-84436-P from Ministerio de Economia y Competitividad, by  Xunta de Galicia ED431C 2017/07 and {\em Centro Singular de Investigaci\'on de Galicia} accreditation 2019-2022, by FEDER and the Spanish Research State Agency by {\em Mar\'\i a de Maeztu Unit of Excellence} MDM-2016-0692.

\bibliographystyle{unsrt}
\bibliography{references}
\end{document}